\documentclass[11pt]{report}
\usepackage{graphics}
\voffset - 1.0in
\title{THE MUON ANOMALOUS MAGNETIC MOMENT\\ - SIGNIFICANCE OF THE NEW
MEASUREMENT }
\author{JACK L. URETSKY \\ High Energy Physics Division,
Argonne National Laboratories}
\date{\today}
\begin{document}
\maketitle
\section{INTRODUCTION}
\indent The beautiful experiment measuring the muon anomalous magnetic moment
(g-2) now in progress at Brookhaven \cite{Brown} has already attained a
precision that challenges our ability to calculate the expected result of the
experiment.  The uncertainty of the Brookhaven g-2 measurement is now about 1.5
in units of $10^{10}a_{\mu}$ ( where $a_{\mu}=\frac{g-2}{2}$), with 1 year of
data yet to be analyzed according to reference 1.  Weak and electromagnetic
contributions to g-2 can be calculated according to a well-understood theory
with about 1/30'th of the uncertainty of the measurement\cite{Mel}, as is noted
in the reference.  The difference between that calculation and the measurement
is 71.8 with the uncertainty of the experimental result.  The difference is
ascribed to hadronic effects (see \cite{Hughes} for a recent review).

\indent The significance of the new result is, in my view, that it measures the
hadronic contribution to the muon g-2 with a precision of the order of 1 \%.
The precision may increase by about an order of magnitude when the experiment is
completed.  There is no theory of hadrons that predicts hadronic interactions so
precisely.  The appropriate theory should presumably be grounded in quantum
chromodynamics (QCD), which does not yet lend itself to such precise
predictions.  There are phenomenological arguments which relate the hadronic
contribution to g-2 to other experiments and provide the basis for current
estimates which range \cite{Yndur} from 69.2 to 72.5, each with a quoted
uncertainty of the order of a percent.

\indent  The quoted uncertainties of current estimates reflect only the
uncertainties of the experimental data used for phenomenological estimates and
the computational uncertainties of the estimates.  They do not, in my view,
adequately reflect the uncertainties in the theories underlying the
phenomenological estimates, as I shall explain in the remainder of this note.

\section{DESCRIPTION OF THE HADRONIC CONTRIBUTION}
\indent  In the notation of Sommerfield \cite{CMS}, the muon propagator $G_{mu}$
relates to a mass function $M$ as
\begin{equation}
G_{mu}^{-1} = \gamma \Pi + M
\label{eq:gmu}
\end{equation}
where $\Pi _{\mu} = (p - eA)_{\mu}$, $p$ is the muon momentum and $A$ is the
potential for a weak constant external field.  The anomalous g-factor occurs as
a term proportional to $\sigma_{\mu \nu}F_{\mu \nu}$ in the mass function $M$.
That function is calculated from the equation
\begin{equation}
M = m_{0} + ie^{2}\gamma G_{mu}\Gamma \cal G
\label{eq:M}
\end{equation}
involving the photon Green's function $\cal G$ and the ``dressed photon-muon
vertex '' $\Gamma $.

\indent The photon Green's function is assumed \cite{Durand} to obey a
K\"{a}llen-Lehman representation \cite{Lehman}.   According to that
representation each element $\cal G_{\mu,\nu}$ may be written as a dispersion
integral
\begin{equation}
{\cal  G}_{\mu,\nu}(k^{2})=\frac{\delta _{\mu,\nu}}{k^{2}} - [\delta _{\mu,\nu}
- \frac{k_{\mu}k_{\nu}}{k^{2}}]\int_{0}^{\infty}\frac{s(x)}{x+k^{2}-
i\epsilon}dx.
\label{eq:D1}
\end{equation}
The weight function $s(x)$ is proportional to the sum over possible intermediate
states produced by the electromagnetic current, according to the equation
\begin{equation}
k^{4}s(-k^{2})\delta _{\mu \nu}
=\frac{1}{3}\sum_{\alpha}<0\mid j_{\mu}(0)\mid \alpha><\alpha \mid
j_{\nu}(0)\mid 0>\delta ^{4}(\alpha - k)
\label{eq:D2}
\end{equation}
where $j$ is the electromagnetic current operator. The hadronic contribution to
$\cal G$ is then given by that part of $s(x)$ for which the states $\mid
\alpha>$ consist of hadrons.  The matrix elements $<0\mid j_{\mu}\mid \alpha >$
are the amplitudes for production of the state $\mid \alpha >$ from the vacuum
by the electromagnetic current.  The corresponding contributions to $s(x)$ are
therefore proportional, to lowest order in the fine-structure constant, to the
cross-sections for producing states $\alpha$ in $e^{+}e^{-}$ collisions.  It
would accordingly seem that such measured cross-sections could be used to
calculate the hadronic contributions to $\cal G$.

\indent  The components $\Gamma_{\mu}$ of the vertex function $\Gamma$ contain
terms corresponding to the production of a virtual hadron that couples to the
muon.  The lowest order such terms (in powers
of the charge $e$) - the so-called ``light-light scattering'' terms - are of
order $e^{6}$.  Power counting suggests that such terms contribute to the
anomalous muon moment at a level of about $\frac{1}{10}$th of the uncertainty in
the Brookhaven measurement.  There is, however, no theory comparable to
electroweak theory for calculating these terms, as Hayakawa and Kinoshita have
emphasized \cite{HayKin} in connection with their own estimates.  Those
estimates are not severely tested at the present level of experimental accuracy.

\section{UNCERTAINTIES IN HADRONIC-CONTRIBUTION ESTIMATES}
\indent The K\"{a}llen-Lehman representation \cite{Lehman} for the photon
propagator is of the nature of a dispersion relation.  The dispersion relation
is a statement that ${\cal G}_{\mu,\nu}(k^{2})$ is a real analytic function of
the variable $k^{2}$, cut along the negative (that is, timelike) $k^2$ axis
\cite{Kniehl}.  It is also, however, a statement that $\cal G$ is a
polynomially-bounded function for large, complex $k^{2}$.  I contend that the
latter statement is unprovable\cite{Weinberg} \cite{Oehme}.  There does not, in any event,
appear to be any existing proof that $\cal G$ obeys such a boundedness property.
Any attempt to assign a numerical uncertainty to a calculation based upon a K\"{a}llen-Lehman representation for the photon propagator must therefore be
regarded as speculative.

\indent There are recent estimates of the hadronic contributions to the muon g-2
that are within a few per cent of values that one would deduce from the new
experimental result \cite{Brown}.  Is this near agreement evidence for the
polynomial boundedness of $\cal G$?  I present in the next section models, by
way of counter-examples, where such near agreement can occur even with functions
that grow exponentially.  Polynomial boundedness cannot, therefore, be deduced
from the present data.  Uncertainty estimates based upon dispersion-relation
calculations of $\cal G$ must be understood as lower limits because the theories
underlying the calculations are incompletely defined.

\section{Dispersion Relations for Functions Having Exponential Growth}
\indent I present two models to show that an attempt to calculate a function by
a dispersion relation, when the assumption of polynomial boundedness is
incorrect, need not lead to a result that is wildly different from the correct
result.  The fact that dispersive calculations of the hadronic component of the
muon g-2 are close to the values deduced from experiment is, therefore, not
evidence that such calculations can give correct results.

\indent Both of the models happen to involve functions related to Bessel
functions.  The essence of the demonstration, however,  may be made with the two    elementary functions natural logarithm and exponential.  Consider, for this purpose
 the following function $Q(z)$ of a complex variable $z$:
\begin{equation}
Q(z) = \ln (-z) + e^{-bz}
\label{eq:Q}
\end{equation}
where $b$ is some real positive constant.  $Q(z)$ is an analytic function
of the complex variable $z$ except for logarithmic branch points at $0$
and $\infty$ and an essential singularity (of the exponential) at $\infty$.
$Q$ is made single-valued by a cut along the positive real axis, giving $Q$
an imaginary part equal to $-i\pi$ as the real axis is approached from above.

The point of the models is that they involve functions that, like $Q$,
have an essential singularity and are therefore not polynomially bounded.
One may nevertheless try write a dispersion relation for $Q$ by integrating
around the cut, because the imaginary part, being a constant in this case,
is well-behaved.  The result of the dispersion integral will be to recover
only the $\ln$ part of $Q$.  The remaining part of $Q$, the exponential,
will be negligible (compared with the logarithm) almost everywhere along the
positive real axis, depending upon the size of the parameter $b$.
The erroneous assumption that $Q$ satisfies a dispersion relation therefore
makes only a small, model-dependent error almost everywhere along the
positive real axis.

\subsection{A Hankel-Function Model}
\indent In the first model I assume that the correct function that I am trying
to calculate is the Hankel function\cite{Abram} $H_{0}^{(1)}(z)$.  The imaginary
part of the Hankel function is the Neumann function $Y_{0}(z)$.  Both functions
emulate a coulomb-like singularity at $z=0$, albeit only logarithmically, so a
numerical integration of a dispersion relation must be cut off at small $z$.
The function $H_{0}^{(1)}(z)$ behaves for large, complex $z$ like
$\frac{e^{iz}}{\sqrt{z}}$, and is not, therefore, polynomially bounded.

\indent The model is unrealistic in the sense that the imaginary part of the
``true function'' is oscillatory, unlike the strictly positive cross-section
data that is used as input in the photon-propagator calculations.  That
particular unrealistic feature is corrected in the second model.

\indent The game, then, is to pretend that we know from experiment the imaginary
part of the ``correct'' function.  We then put the ``experimental'' data into an
unsubtracted dispersion relation under the erroneous assumption that the
``correct'' function vanishes sufficiently rapidly at large $|z|$, cutting off
the integration at some small value of $z$.  We want to determine how much our
answer deviates from the correct answer which is, in this case, the Bessel
function $J_{0}(z)$.

\indent The answer can be determined analytically, because
\begin{figure}
\scalebox{.75}
\centering \includegraphics{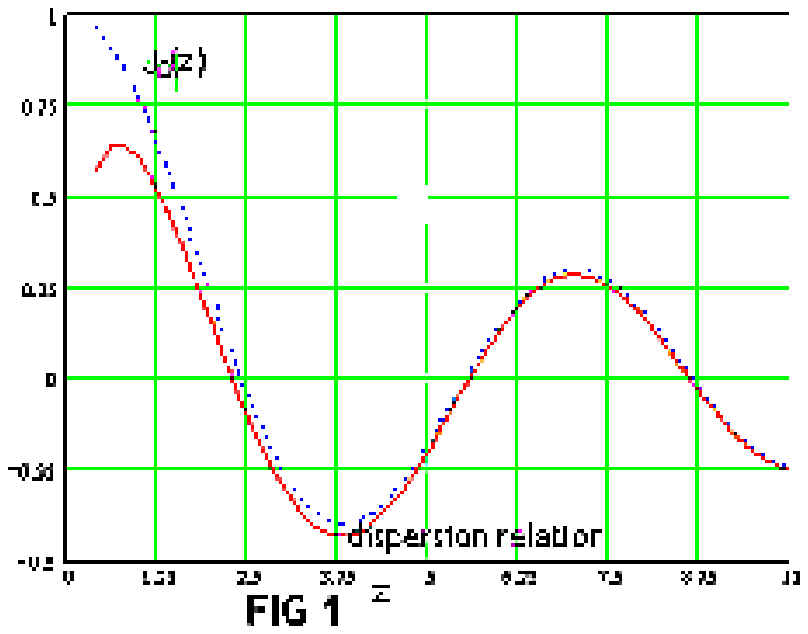}
\end{figure}
\begin{equation}
\frac{1}{\pi}\int_{0}^{\infty}\frac{Y_{0}(x)dx}{x-z}=J_{0}(z)-
\frac{2}{\pi^{2}}S_{-1,0}(z)
\label{eq:J}
\end{equation}
where the principal value of the integral is intended and $S_{-1,0}(z)$ is a
Lommel function whose explicit form has been given by Watson \cite{Wat}.  It is
a solution of the inhomogeneous Bessel Equation
\begin{equation}
S^{\prime \prime} + \frac{1}{z}S^{\prime} + S = \frac{1}{z^{2}}
\label{eq:Lom}
\end{equation}
subject to the condition that it behaves asymptotically like $\frac{1}{z^{2}}$.
It grows for small $z$ like $ln^{2}(z)$, but rapidly becomes small as $z$
increases.  Fig. 1 compares the right hand side of Eq. \ref{eq:J}, shown by the
solid line, with the ``correct'' answer $J_{0}(z)$, shown by the dotted line.
It is apparent that the error from using misusing the dispersion integral can be
made quite small if small values of $z$ are excluded, as they are in ``real
life'' when coulomb and infra-red effects play a role.
\subsection{A Strictly Positive Imaginary Part}
\indent A more realistic model would require that the simulated experimental
data be given by a function that is strictly positive, like a cross-section.
A convenient choice is the modified Hankel function $K_{0}(z)$ which
approaches zero asymptotically for large $z$ like $\frac{e^{-z}}{\sqrt{z}}$
and grows logarithmically for small $z$ like $-ln(z)$.  For $z$
approaching the positive real axis from above, there is a function $f(z)$
given by
\begin{equation}
f(z)= \frac{1}{\pi}\int_{0}^{\infty}K_{0}(x)dx(\frac{1}{x-z}-\frac{1}{x+z}) =
\pi ({\bf L_{0}}(z)-I_{0}(z)) + iK_{0}(z)
\label{eq:K1}
\end{equation}
Here $\bf L_{0}$ is the modified Struve function and $I_{0}$ is the modified
Bessel function, each of order zero.

\indent The complex function $f(z)$ does not satisfy a dispersion relation.
This is because the integral of the function around a closed loop at large $z$
cannot be made arbitrarily small, since its imaginary part, $K_{0}(z)$, grows
without bound when $\frac{\pi}{2}< arg(z)<\frac{3\pi}{2}$.  The imaginary part
of $f(z)$ again represents the simulated experimental data.

\begin{figure}
\scalebox{.75}
\centering \includegraphics{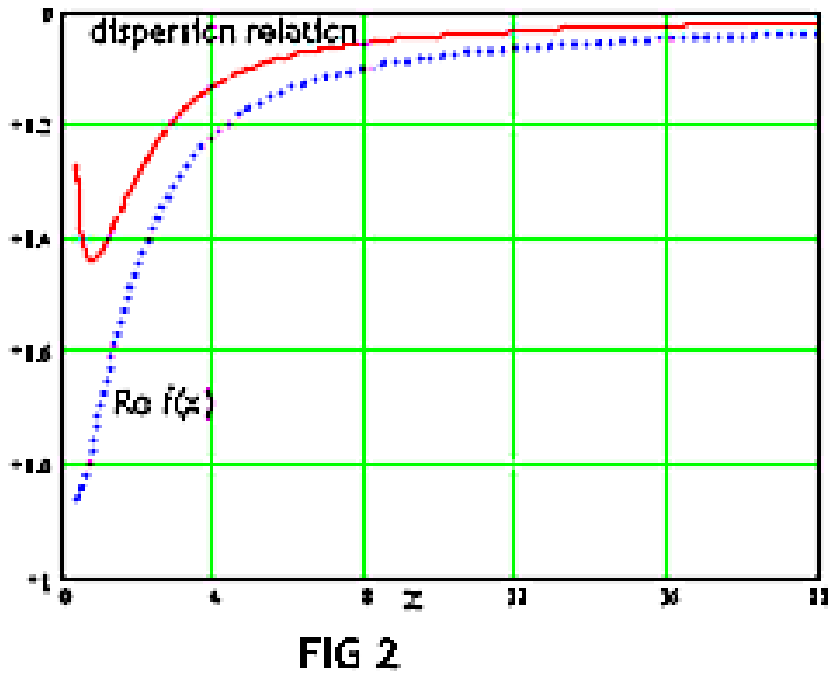}
\end{figure}
\indent The game, as before, is to put the simulated experimental data into an
unsubtracted dispersion relation and compare with the correct answer, given in
Eq \ref{eq:K1}.  The result is shown in Fig. 2 which displays the real part of
the dispersion result as a solid line and the real part of $f(s)$ as a dotted
line (the imaginary parts are, of course, identical).  It is evident that for
values of $z$ greater than about unity, the dispersion result is not greatly
different from the ``correct'' value.
\section{Conclusion}
\indent Hadronic contributions to g-2 originate, according to our present
understanding, from electromagnetic excitation of virtual quark pairs from the
vacuum.  The task of calculating the effects of such excitations therefore falls
within the realm of Quantum Chromodynamics (``QCD'').  Such calculations have
not yet been done.  It has been tempting to suppose, as an alternative, that
dispersion relations using empirical data could be used to predict the results
of QCD calculations, at least for the vacuum polarization corrections to the
photon propagator.

\indent The difficulty with the supposition is that a dispersion relation
involves an assumption about the behavior of a function for large values of its
argument, namely, that the function is at least polynomially bounded.  In the
case of the photon propagator there is no way to justify that assumption.

\indent It might be tempting to try to evade this lack of justification by reversing
the burden of proof, asking: "What is the physical origin of the analogous
terms in the physically relevant case?  In particular, if
Exp[-z] ---$>$  Exp[-E/E0], what is the meaning and value of the
energy scale E0?"

\indent Because the presence of an essential singularity at infinite
energy, as I have modeled with an exponential term, is purely conjectural,
any mass, such as the pion mass, the QCD scale or a spontaneously generated
mass \cite{Wu} is a possible candidate for the energy scale E0.  One doesn't
see such
terms in the usual discussions of analytic properties of amplitudes because those discussions are customarily guided by perturbation theory \cite{Eden} \cite{Oehme}, where such terms cannot occur.  But we know of models where perturbation theory breaks down. \cite{Wu}.  And essential singularities are a common feature of solutions of differential equations such as the relativistic Schroedinger equation, {\em cf.} \cite{Olsson}.

\indent Proof of the existence of an essential singularity would, of course,
end the discussion; dispersion relations would not provide an algorithm
for
calculating the hadronic contribution to vacuum polarization.  That is not
the intent of this note, however.  The intent is merely to point out that
there is,
as yet, no mathematical justification for the use of dispersion relations
for that purpose, and the accuracy of the resulting predictions must
therefore remain uncertain.

\indent  What I have shown here is that violation of the assumption of
polynomial boundedness can, in some circumstances, lead to results that are
close to the correct results.  There is, however, no apparent {\em a priori}
way to estimate the size of the error resulting from the violation.  There is
consequently no way to estimate the error involved in using a dispersion
relation to calculate the hadronic contribution in question.  \cite{Uretsky}

\indent  The Brown, {\em et al.}\cite{Brown} measurement of the muon g-2 is
therefore significant because it measures the hadronic contribution to g-2 with
unprecedented precision.  That measurement stands as a challenge to our
understanding of the quark structure of the vacuum. \cite{Uretsky}

\indent  I am indebted to Geoffrey Bodwin, Tim Tait, Carlos Wagner,
Alan White, and Cosmas Zachos for some helpful discussions,
to Professor T. T. Wu for his assistance in identifying reference
\cite{Wu}, and to the (anonymous) referee, whose
question led me to become aware of references \cite{Oehme}, \cite{Wu},
\cite{Olsson} and several related papers.  This work was supported by the
U.S. Department of Energy, Division of High Energy Physics,
Contract W-31-109-ENG-38.


\begin{thebibliography}{99}
\bibitem{Brown}Brown, H. {\em et  al.} Phys. Rev. Lett. {\bf 86}, 2227 (2001).
\bibitem{Mel}But see K. Melnikov, hep-ph/0105267, which was posted during the
writing of the present note.
\bibitem{Hughes} Hughes, V.W. {\em Reviews of Modern Physics}, 71 (1999) S133
\bibitem{Yndur}Yndur\'{a}in, http//xxx.lanl.gov/abs/ hep-ph/0102312
\bibitem{CMS} C. M. Sommerfield, {\em Ann. Phys.} (N.Y.) 5 (1958) 26
\bibitem{Durand} C Bouchiat and L. Michel, J. Phys. Radium 174 (1961) 1835,
L. Durand III, Phys. Rev. 128 (1962) 441, followed in, {\em e.g.}, R.
Alemany, M. Davier and A. Hocker, European Physical Journal C 2 (1998) 111.
\bibitem{Lehman} G. K\"{a}llen, Helv. Phys. Acta {\bf 25} (1952 417, H.
Lehman, Nuovo Cimento {\bf 11}(1954) 342.
\bibitem{HayKin}M. Hayakawa and T. Kinoshita, Phys. Rev. D \underline{57} (1998)
465
\bibitem{Kniehl}A recent review is Bernd A. Knield's lecture {\em XXXVI Cracow
School of Theoretical Physics}, hep-ph/9607255.
\bibitem{Weinberg}See, in this connection, S. Weinberg {\em The Quantum Theory
of Fields} (Cambridge University Press 1995), Vol. I, p. 460 (fn.).
\bibitem{Oehme} Oehme and Zimmermann, Phys. Rev. \underline{22} (1980) 2534, discuss assumptions about properties of QCD amplitudes that would make a proof of polynomial boundedness possible.  Their discussion relies on QCD perturbation theory.
\bibitem{Abram} I use the notation of Abramowitz and Stegun, {\em Handbook of
Mathematical Functions} (Dover Publications, New York 1972)
\bibitem{Wat}G.N. Watson, {\em Theory of Bessel Functions} (Cambridge U. Press
1948), p. 349
\bibitem{Wu} As recognized in footnote 14 of reference \cite{Oehme}, citing McCoy and Wu, [MPI Report No. MPI-PAE/PTh 35/79, Munich, 1979] published as Phys. Lett. \underline{87B} (1979) 50.
\bibitem{Eden} R. J. Eden, {\em et al.}, {\em The Analytic S-Matrix} (Cambridge University Press 1966), see especially pp. 3-5
\bibitem{Olsson} M. G. Olsson, Phys. Rev. D \underline{56} (1997) 238
\bibitem{Uretsky} See my letter and Prof. Lee Roberts' response in
Physics World \underline{14} {June 2001} 23
\end{thebibliography}
\end{document}